\documentclass[aps, prd, twocolumn,print, showpacs, nofootinbib]{revtex4}
\usepackage{txfonts}
\usepackage{amssymb}
\usepackage{mathrsfs}
\usepackage{graphicx}
\usepackage{epsfig}
\usepackage{dcolumn}
\usepackage{bm}

\begin{document}

\title{Bondi accretion of dark matter by neutron stars}

\author{Xi Huang,$^{1,3,}$\footnote{huangxi@mails.ccnu.edu.cn} Jian-Feng Liu,$^2$ Wei-Hua Wang,$^2$ Quan Cheng,$^4$ and Xiao-Ping Zheng$^{2,}$\footnote{zhxp@phy.ccnu.edu.cn}}

\address{$^{1}$School of Electronic and Electrical Engineering, Wuhan Textile University, Wuhan 430073, China\\
$^{2}$Institute of Astrophysics, Central China Normal University, Wuhan 430079, China\\
$^{3}$Key Laboratory of Quark and Lepton Physics (Ministry of Education), Central China Normal University, Wuhan 430079, China\\
$^{4}$Key Laboratory of Particle Astrophysics, Institute of High
Energy Physics, Chinese Academy of Sciences, Beijing 100049,
China\\}



\begin{abstract}

In this paper, we have compared two different accretion mechanisms
of dark matter particles by a canonical neutron star with
$M=1.4~M_{\odot}$ and $R=10~{\rm km}$, and shown the effects of dark
matter heating on the surface temperature of star. We should take
into account the Bondi accretion of dark matter by neutron stars
rather than the accretion mechanism of Kouvaris (2008)
\citep{Kouvaris08}, once the dark matter density is higher than
$\sim3.81~\rm GeV/cm^3$. Based on the Bondi accretion mechanism and
heating of dark matter annihilation, the surface temperature
platform of star can appear at $\sim 10^{6.5}$ year and arrive $\sim
1.12\times10^5$ K for the dark matter density of $3.81~\rm
GeV/cm^3$, which is one order of magnitude higher than the case of
Kouvaris (2008) with dark matter density of $30~\rm GeV/cm^3$.

\end{abstract}

\maketitle

\section{INTRODUCTION} \label{intro}

According to Planck measurements of the cosmic microwave background
(CMB) temperature and lensing-potential power spectra \citep{Ade16},
we have known that the total energy of the Universe contains about
27\% dark matter (DM) and about 5\% baryon matter as well as about
68\% dark energy. It is generally believed that DM particles only
participate in gravitational interaction, but not in electromagnetic
and strong interaction. Therefore, it is very difficult to detect DM
particles. In recent years, there are mainly three kinds of methods
to detect DM particles, namely, the direct detections, the indirect
detections, and the collider detections \citep{Bi13,Klasen15}.
Direct search experiments aim to detect the optical, thermal, and
ionized signals due to the scattering between DM particles and
target nuclei of detectors \citep{Goodman85}, such as CRESST-II
\citep{Angloher16}, SuperCDMS \citep{Agnese16}, DarkSide
\citep{Agnes16}, LUX \citep{Akerib17}, XENON100 \citep{Aprile16},
PandaX-II \citep{Tan16} experiments and so on. Indirect detection
experiments mainly probe the signals that DM particles can decay or
annihilate into gamma-rays, neutrinos or charged anti-particles such
as positrons and antiprotons \citep{Bertone05}, including AMS-02,
PAMELA, Fermi-LAT, IACTs, IceCube, ANTARES, Super-K experiments
\citep{Rott13,Conrad14}, etc. Finally, the collider detection is to
produce DM particles via the high energy collider, such as the Large
Hadron Collider (LHC) at CERN \citep{Aad15,Khachatryan16}. From a
theoretical point of view, DM is stable in the cosmic age scale,
lightless, and massive particle beyond the Standard Model of
particle physics. To date, several candidates of DM particles are
supposed \citep{Baer15}, for example, weakly interacting massive
particle (WIMP) such as Kluza-Klein particle \citep{Hooper07}
predicted by the theories of the additional spatial dimension and
neutralino \citep{Jungman96} proposed in the supersymmetric
theories, axion \citep{Weinberg78,Wilczek78}, axino \citep{Covi01},
gravitino \citep{Pagels82}, millicharged (MC) particle
\citep{Goldberg86}, self-interacting DM (SIDM) \citep{Carlson92} and
so on.

At first, in order to solve the neutrino problem in the Sun and the
missing mass problem in the Galaxy, Press and Spergel (1985)
\citep{Press85} studied that WIMPs are captured by the Sun via the
captured mechanism of scattering off individual nucleons in the Sun
and into bound orbits. Subsequently, Gould (1987,1988)
\citep{Gould87,Gould88} computed the direct and indirect capture of
WIMPs by the Earth. It is interesting that Goldman and Nussinov
(1989) \citep{Goldman89} researched that WIMPs are trapped in
neutron stars and concentrated towards the star center. This results
in the formation of a mini black hole that consumes the neutron star
(NS). Recently, basing on the work of Press and Spergel (1985),
Kouvaris (2008) \citep{Kouvaris08} discussed in detail that WIMPs
are accreted and trapped eventually by the NS, taking into account
the general relativity corrections of the star. Meanwhile, the
WIMP's accretion rate onto the NS was shown, and the effect of the
WIMP annihilation on the cooling curves of a canonical NS made of
ordinary nuclear matter was calculated. The main conclusions of this
paper \citep{Kouvaris08} are that the released energy due to WIMP
annihilation inside the NS might affect the temperature of the star
older than $10^7$ years, appearing the plateau of surface
temperature at $\sim 10^4 \rm~K$ for a typical NS. By the way,
besides the above-mentioned heating effects owing to the
annihilation of trapped DM particles onto the NS
\citep{Kouvaris08,Bertone08,Kouvaris10,de Lavallaz10,Fan11,Huang14},
the rotation effect on the NS was also studied in our previous work
\citep{Huang15}, based on an enhanced slow-down of neutron stars due
to an extra current yield by the accretions of millicharged dark
matter particles (MCDM) \citep{Kouvaris14}. We constrained the
charge-mass phase space of MC particles through the simulation of
the rotational evolution of neutron stars \citep{Huang15}.

In this paper, we revisit the accretion mechanism of DM particles by
neutron stars basing on the work of Kouvaris (2008)
\citep{Kouvaris08}. We should consider the Bondi accretion
\citep{Bondi52,Kato98, Edgar04} of the SIDM rather than the
accretion mechanism of Kouvaris (2008), if the candidate of DM is in
the form of the SIDM
\citep{Carlson92,David00,Kaplinghat14,Kamada16,Robertson17}, and the
DM density is about one order of magnitude higher than the standard
DM density 0.3 $\rm GeV/cm^3$ around the Earth. As discussed in a
large amount of literatures, the SIDM can affect the mass profile
and shape of DM halos, reducing the central densities of dwarfs and
low surface brightness galaxies in agreement with observations
\citep{Kaplinghat14}. It is very likely that the SIDM is made up of
composite states of baryons or pions in the dark sector. The most
attractive characteristic of SIDM is large scattering cross section
with other DM particles. The commonly accepted view is that the
cross section per unit mass of SIDM must be of order
\begin{equation}
\sigma_{X}/m_{X}\sim1~{\rm cm^2/g}\approx2\times10^{-24}~{\rm
cm^2/GeV}
\end{equation}
or larger, where $X$ is the DM particle. Obviously, this value is
many orders of magnitude larger than that of the WIMP. The typical
cross section of WIMP is $\sigma_{X}\sim10^{-36}~{\rm cm^2}$, if the
mass is $m_{X}\sim100~{\rm GeV}$, thus the cross section per unit
mass is $\sigma_{X}/m_{X}\sim10^{-38}~{\rm cm^2/GeV}$. For the SIDM,
we can easily deduce the DM density
$\rho=\frac{1}{\sqrt{2}\left(\sigma_{X}/m_{X}
\right)\overline{\lambda}}\approx3.81~\rm GeV/cm^3$, if the mean
free path takes the diameter of Milky Way galaxy
$\overline{\lambda}\sim30~{\rm kpc}$. We consider there is no doubt
that the Bondi accretion of DM by neutron stars is dominant as long
as the DM density is higher than the above value.

This paper is organized as follows. First, the thermal evolution
equations of a canonical NS ($M=1.4~{\rm M}_{\odot}$, $R=10$ km) and
the accretion rate due to the Bondi accretion of DM particles, are
shown in Sec. II. Then we numerically calculate the effect of SIDM
annihilation on the surface temperature of the NS compared to the
work of Kouvaris (2008) in Sec. III. Finally, we present the
conclusions in Sec. IV.

\section{THE MODELS} \label{model}

\subsection{The thermal evolution equations of the NS}

In general, for a newborn NS, the internal temperature can reach up
to $10^{10}$ K, it cools via neutrino emission during the first
million years (neutrino cooling era). Incidentally, the contribution
to neutrino emission would mainly come from the interior of the NS,
without considering it from the crust. The emission of photons from
the surface of the star dominates the cooling (photo cooling era),
as soon as the internal temperature drops below $10^{8}$ K. Besides
the NS cools via the internal neutrino emission and surface photon
emission, on the other hand, various heating mechanisms can be
present during the late times of the thermal evolution
\citep{Schaab99,Gonzalez10}. These include the frictional
interaction between the faster rotating superfluid core and the
slower rotating outer solid crust \citep{Shibazaki89,Larson99},
crust cracking \citep{Cheng92}, magnetic field decay
\citep{Goldreich92,Thompson96,Heyl98,Pons07}, rotochemical heating
\citep{Reisenegger95,Reisenegger97,Fer05}, deconfinement heating
\citep{Yuan99,Yu06}, and dark matter heating \citep{Kouvaris08,de
Lavallaz10,Huang14}.

Direct Urca (DU) processes and modified Urca (MU) processes are
dominant for the neutrino emission. DU processes $n\rightarrow
p+e+\bar{\nu}_e$, $p+e\rightarrow n+\nu_e$ are allowed in
sufficiently dense nuclear matter, nuclear matter with pion
condensation, kaon condensation, or nonzero hyperon density, and in
all phases of quark matter except the Color Flavor Locked phase
\citep{Alford05}. The temperature of the star drops very fast via
the DU processes, in which the emissivity scales as
$\epsilon_\nu\sim T^6$ ($T$ is the interior temperature of the
star). However, DU processes are kinematically forbidden if the star
is not in sufficiently high density. In this case, MU processes
$b+n\rightarrow b+p+e+\bar{\nu}_e$, $b+p+e\rightarrow b+n+\nu_e$ are
switched on in NS cores, which $b$ denotes the bystander (neutron or
proton). The neutrino emissivity of MU processes can be written as
\citep{Shapiro83}
\begin{equation}
\epsilon_{\nu}=1.2\times10^{4} \left({\frac{n_{\rm b}}{{
n}_0}}\right)^{2/3}\left(\frac{T}{10^7~ \rm K}\right)^8 {\rm
\hspace{0.1cm} ergs \hspace{0.1cm} cm^{-3} \hspace{0.1cm} s^{-1}},
\end{equation}
where $n_{\rm b}$ is the average baryon number density, $n_{\rm
0}=0.17{\rm \hspace{0.1cm} fm^{-3}}$ is the nuclear saturation
density.

After the first million years, the dominant mechanism of cooling is
the photon emission from the surface of the star. The surface photon
luminosity of the star is $L_{\rm \gamma}=4\pi R^{2}\sigma T_{\rm
S}^{4}$, where $\sigma$ is the Stefan-Boltzmann constant and $T_{\rm
S}$ is the surface temperature of the star. The relationship between
the interior temperature $T$ and the surface temperature $T_{\rm S}$
is well approximated by \citep{Gudmundsson83}
\begin{equation}
T_{\rm S}=(0.87\times 10^6~{\rm K})\left(\frac{g_{\rm s}}{10^{14}~
{\rm {cm/s}^2}}\right)^{1/4} \left( \frac{T}{10^8~{\rm {K}}}
\right)^{0.55},
\end{equation}
where $g_{\rm s}=GM/R^2$ is the surface gravity of the star. Thus,
the surface photon luminosity $L_{\rm \gamma}$ can be expressed in
terms of the interior temperature as
\begin{equation}
L_{\rm \gamma}=4\pi R^2\sigma(0.87\times 10^6~{\rm
K})^4\left(\frac{g_{\rm s}}{10^{14}~ {\rm {cm/s}^2}}\right) \left(
\frac{T}{10^8~{\rm {K}}} \right)^{2.2}.
\end{equation}
The emissivity from the surface photon emission for the NS of
$M=1.4~{\rm M}_{\odot}$ and $R=10$ km is given by
\begin{equation}
\epsilon_{\rm \gamma}=\frac {L_{\rm \gamma}}{\left({4/3}\right)\pi
R^{3}}=1.8\times10^{14}\left( \frac{T}{10^8~{\rm {K}}}
\right)^{2.2}{\rm \hspace{0.1cm} ergs \hspace{0.1cm} cm^{-3}
\hspace{0.1cm} s^{-1}}.
\end{equation}

It is well known that the NS is a compact object, thus DM around the
star can be accreted. As a result of scattering with nuclei, DM can
lose enough energy and get deposited in the core of the star fast.
It releases a huge amount of energy to heat up the star due to the
annihilation between DM particles. Because of the competition
between the accretion and annihilation of DM, the population of DM
inside the star is governed by
\begin{equation}
\frac{dN}{dt}=\mathcal {F}-\Gamma_{\rm annih}, \label{population}
\end{equation}
where $\mathcal {F}$ (number of particles per time) represents the
accretion rate of DM captured by neutron stars and $\Gamma_{\rm
annih}=\langle \sigma_{\rm annih}\upsilon \rangle \int n^2_X {\rm
d}V=C_A N(t)^2$ represents the annihilation rate. In the above
equation, $\langle \sigma_{\rm annih}\upsilon \rangle$ is the
thermally averaged annihilation cross section times the DM velocity,
$n_X$ is the number density of DM inside the star, and $C_A=\langle
\sigma_{\rm annih}\upsilon \rangle /V$ is the constant where $V$ is
the volume of the star.

Obviously, the solution of Eq. (\ref{population}) is
\begin{equation}
N(t)=\sqrt{\frac{\mathcal {F}}{C_A}}\tanh(t/\tau),
\end{equation}
where $\tau=1/{\sqrt{\mathcal {F}{C_A}}}$ is the time scale,
depending on the DM density and temperature of the star
\citep{Kouvaris08}. Therefore, the luminosity due to the
annihilation of DM particles can be expressed as
\begin{equation}
L_X=\Gamma_{\rm annih}m_X=C_A N(t)^2m_X=\mathcal
{F}\tanh^2(t/\tau)m_{X}. \label{energy}
\end{equation}
As it can be seen from above Eq. (\ref{energy}), the annihilation
rate saturates to the accretion rate when $t$ is larger than roughly
3$\tau$. Hence, the emissivity of DM can be written as
\begin{equation}
\epsilon_X=\frac{L_X}{4 \pi R^3/3}.
\end{equation}

In order to obtain the curves of thermal evolution for a NS, we also
need to know the specific heat of the star. For convenience, the
core of NS is made of non-superfluid neutrons, protons, and few
electrons in our model. Meanwhile, we suppose that the density of
the star is uniform. The specific heat of the star mainly
contributed by the fermions from the core is given by
\citep{Shapiro83}
\begin{equation}
c_V=\frac{k^2_{\rm B}T}{3\hbar^3c}\sum_ip^i_{\rm
F}\sqrt{m^2_ic^2+(p^i_{\rm F})^2},
\end{equation}
where $k_{\rm B}$, $\hbar$, and $c$ are the Boltzmann constant,
reduced Planck constant, and speed of light, respectively, the sum
runs over the different particles $n$, $p$, $e$ and their
corresponding Fermi momenta are
\begin{equation}
p^n_{\rm F}=(340~\rm{MeV})\left(\frac{{\it n}_{\rm b}}{{\it
n}_0}\right)^{1/3},
\end{equation}
\begin{equation}
p^p_{\rm F}=p^e_{\rm F}=(60~\rm{MeV})\left(\frac{{\it n}_{\rm
b}}{{\it n}_0}\right)^{2/3}.
\end{equation}
However, we should emphasize that the contribution to the specific
heat from DM particles can be neglected, because they consist of a
tiny fraction of the whole mass of the star.

By far, we can gain the thermal evolution equation of the NS
\begin{equation}
\frac{dT}{dt}=\frac{-\epsilon_{\nu}-\epsilon_{\gamma}+\epsilon_X}{
c_V}.
\end{equation}

\subsection{The accretion rate of Bondi accretion}

In this section we calculate the Bondi accretion rate of DM
particles along the lines of \citep{Yang92}. As we know the
continuity equation of magneto-fluid is
\begin{equation}
\frac{\partial \rho}{\partial t}+\nabla \cdot(\rho \bm{\upsilon})=0,
\label{continuity}
\end{equation}
where $\rho$ and $\bm{\upsilon}$ are the density and velocity of
fluid, respectively. In the polar coordinate of spherical symmetry,
we have
\begin{equation}
\nabla \cdot(\rho \bm{\upsilon})=\frac{1}{r^2}\frac{\partial
(r^2\upsilon_r\rho)}{\partial r}.
\end{equation}
Thus Eq. (\ref{continuity}) can be written as
\begin{equation}
\frac{\partial \rho}{\partial t}+\frac{1}{r^2}\frac{\partial
(r^2\upsilon_r\rho)}{\partial r}=0.\label{continuity2}
\end{equation}
We can derive $\frac{\partial \rho}{\partial t}=0$, $\frac{\partial
\upsilon_r}{\partial t}=0$, and $r^2\upsilon_r\rho=\rm {const}$
easily, if the fluid flows smoothly. In general, this const is
defined as $\frac{\dot{M}}{4\pi}$, where $\dot{M}$ is the accretion
rate (accreted mass per time). In addition, the momentum equation of
magneto-fluid is
\begin{equation}
\frac{\partial {\bm \upsilon}}{\partial t}+(\bm
\upsilon\cdot\nabla)\bm \upsilon=-\frac{1}{\rho}\nabla p+\nabla
\psi,\label{momentum}
\end{equation}
where $(\bm \upsilon\cdot\nabla)\bm
\upsilon=\upsilon_r\frac{\partial}{\partial r} \upsilon_r$ in the
polar coordinate of spherical symmetry, $p$ is the pressure of
fluid, and $\nabla \psi=-\frac{GM}{r^2}$. Therefore Eq.
(\ref{momentum}) can be simplified as
\begin{equation}
\frac{\partial \upsilon_r}{\partial
t}+\upsilon_r\frac{\partial}{\partial r}
\upsilon_r=-\frac{1}{\rho}\frac{\partial p}{\partial
r}-\frac{GM}{r^2}.\label{momentum2}
\end{equation}

For conveniences, the equation of state of fluid is supposed in the
form of polytropic process, namely, $p\sim \rho^\gamma$. Thus we can
define the acoustic velocity $a_s^2=\frac{\partial p}{\partial
\rho}=\gamma \frac{p}{\rho}$. For Eq. (\ref{momentum2}), we can
deduce
\begin{equation}
\frac{1}{\rho}\frac{\partial p}{\partial
r}=\frac{1}{\rho}\frac{\partial {\rho^\gamma}}{\partial
r}=a_s^2\frac{1}{\rho}\frac{\partial \rho}{\partial r}. \label{p/r}
\end{equation}
According to continuity equation (\ref{continuity2}), we can give
$\frac{\partial}{\partial r}(r^2\upsilon_r\rho)=0$, namely,
\begin{equation}
\frac{1}{\rho}\frac{\partial \rho}{\partial r}=-\frac{1}{\upsilon_r
r^2}\frac{\partial}{\partial r}(\upsilon_r r^2). \label{rho/r}
\end{equation}
Combining Eqs. (\ref{p/r}) and (\ref{rho/r}), Eq. (\ref{momentum2})
can be written as
\begin{equation}
\frac{1}{2}\left(1-\frac{a_s^2}{\upsilon_r^2}\right)\frac{\partial}{\partial
r}(\upsilon_r^2)=-\frac{GM}{r^2}\left(1-\frac{2a_s^2r}{GM}\right).
\label{eq1}
\end{equation}
As the fluid is close to the central compact object, namely, $r$
decreases continually, $\left(1-\frac{2a_s^2r}{GM}\right)$ will
increase gradually. When this value reaches zero, we can obtain
\begin{equation}
r_s=\frac{GM}{2a_s^2(r_s)},\label{rs}
\end{equation}
where $r_s$ is called the critical radius. If $r=r_s$, we can see
the left hand side of Eq. (\ref{eq1}) must be zero, requiring
\begin{equation}
\upsilon_r^2=a_s^2 \label{vr=as}
\end{equation}
or
\begin{equation}
\frac{d}{dr}(\upsilon_r^2)=0.
\end{equation}

On the basis of laws of thermodynamics, there exists
$dh=\frac{dp}{\rho}$ for the isentropic process, where $h$ is the
enthalpy per unit mass of the fluid. For the polytropic process, the
equation of state is $p=p_0\rho^\gamma$ where $p_0$ is constant.
Hence, we can derive
\begin{equation}
\frac{1}{\rho}\frac{dp}{d\rho}=\gamma p_0\rho^{\gamma-2}. \label{dp}
\end{equation}
Defining $h=h_0\rho^\beta$ where $h_0$ is constant, thus we have
\begin{equation}
\frac{dh}{d\rho}=\beta h_0 \rho^{\beta-1}. \label{dh}
\end{equation}
Defining $\beta=\gamma-1$ and combining Eqs.
(\ref{dp})$-$(\ref{dh}), we can deduce
\begin{equation}
h=\left(\frac{\gamma}{\gamma-1}\right)\frac{p}{\rho}=\frac{a_s^2}{\gamma-1}.
\label{h}
\end{equation}

In addition, since the fluid flows steadily, Eq. (\ref{momentum2})
can be written as
\begin{equation}
\upsilon_r\frac{\partial}{\partial r}
\upsilon_r=-\frac{1}{\rho}\frac{\partial p}{\partial
r}+\frac{\partial \psi}{\partial r}=-\frac{\partial h}{\partial
r}+\frac{\partial \psi}{\partial r},
\end{equation}
namely,
\begin{equation}
\frac{\upsilon_r^2}{2}+h-\psi=B, \label{eq2}
\end{equation}
where $B$ is constant.

In the following section, we will compute the integration constant
$B$ using the boundary conditions. When $r\rightarrow r_\infty$, the
radial velocity of fluid and gravitational potential also approach
to zero $\upsilon_\infty \rightarrow 0$, $\psi_\infty \rightarrow
0$. In terms of Eq. (\ref{eq2}), we can get $h_\infty =B$. Basing on
Eq. (\ref{h}), thus we have
\begin{equation}
h_\infty=\frac{a_{s\infty}^2}{\gamma-1}=B. \label{h-infinity}
\end{equation}
Combining Eqs. (\ref{rs}), (\ref{vr=as}), and (\ref{h}), Eq.
(\ref{eq2}) can be written as
\begin{equation}
\frac{a_s^2(r_s)}{2}+\frac{a_s^2(r_s)}{\gamma-1}-2a_s^2(r_s)=B
\end{equation}
at the critical radius $r=r_s$. In the above equation, the
integration constant $B$ is given by
\begin{equation}
\frac{a_s^2(r_s)(5-3\gamma)}{2(\gamma-1)}=B. \label{B}
\end{equation}

Comparing Eq. (\ref{h-infinity}) to Eq. (\ref{B}), we can easily
derive
\begin{equation}
a_s^2(r_s)=\frac{2a_{s\infty}^2}{5-3\gamma}. \label{as(rs)}
\end{equation}
Hence, the critical radius can be expressed as
\begin{equation}
r_s=\frac{GM}{2a_s^2(r_s)}=\frac{(5-3\gamma)GM}{4a_{s\infty}^2}.
\label{rs2}
\end{equation}
According to the definition of acoustic velocity $a_s^2\sim
\rho^{\gamma-1}$, we have
\begin{equation}
\frac{a_s(r)}{a_{s\infty}}=\left[ \frac{\rho(r)}{\rho(r_\infty)}
\right]^{\frac{\gamma-1}{2}}. \label{as(r)}
\end{equation}
In terms of Eq. (\ref{as(rs)}), Eq. (\ref{as(r)}) can further be
written as
\begin{equation}
\left[\frac{a_s(r_s)}{a_{s\infty}}\right]^2=\left(\frac{2}{5-3\gamma}\right)=\left[\frac{\rho(r_s)}{\rho(r_\infty)}\right]^{\gamma-1}
\end{equation}
at the critical radius $r=r_s$. Therefore, the density of fluid at
the critical radius can be expressed as
\begin{equation}
\rho(r_s)=\rho(r_\infty)\left(\frac{2}{5-3\gamma}\right)^{\frac{1}{\gamma-1}}.
\label{rho(rs)}
\end{equation}
The accretion rate at the critical radius can represent those with
different radii, because it is constant due to flowing steadily.
Finally, combining Eqs. (\ref{vr=as}), (\ref{as(rs)}), (\ref{rs2}),
and (\ref{rho(rs)}), the accretion rate (accreted mass per time) is
given by
\begin{eqnarray}
\dot{M}&=&4\pi\rho(r_s)\upsilon_r(r_s)r_s^2
\nonumber \\
&=&
4\pi\rho(r_\infty)\left(\frac{2}{5-3\gamma}\right)^{\frac{1}{\gamma-1}}\left(\frac{2}{5-3\gamma}\right)^{\frac{1}{2}}a_{s\infty}\left(\frac{5-3\gamma}{4}\right)^2\frac{G^2M^2}{a_{s\infty}^4}
\nonumber \\
&=&\Gamma
\rho(r_\infty)\frac{G^2M^2}{a_{s\infty}^3},\label{accretion rate}
\end{eqnarray}
where
$\Gamma=4\pi\times2^{\frac{1}{\gamma-1}-\frac{7}{2}}(5-3\gamma)^{\frac{5-3\gamma}{2(1-\gamma)}}$.
For the adiabatic process, we know $\gamma=\frac{5}{3}$, thus
$\Gamma=\pi$. According to the definition of acoustic velocity, we
can further calculate the acoustic velocity at infinity
$a_{s\infty}^2=\gamma\frac{\pi}{8}\bar{\upsilon}_\infty^2$. Hence,
Eq. (\ref{accretion rate}) can further be written as
\begin{equation}
\dot{M}=\pi\left(\frac{5\pi}{24}\right)^{-\frac{3}{2}}\rho(r_\infty)\frac{G^2M^2}{\bar{\upsilon}_\infty^3}.\label{accretion
rate 2}
\end{equation}

\section{NUMERICAL RESULTS AND DISCUSSIONS} \label{results}

\begin{figure}
\centering
\includegraphics[scale=0.43]{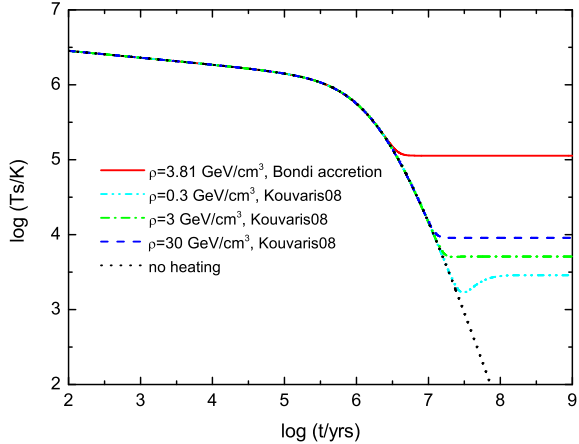}
\caption{The thermal evolution of a canonical NS with
$M=1.4~M_{\odot}$ and $R=10~{\rm km}$. The dotted curve that crosses
the time axis represents the ``standard cooling" case without
considering the DM heating. The dash-dot-dotted, dash-dotted, and
dashed curves from bottom to top correspond to DM densities of 0.3,
3, and 30 $\rm{GeV/cm^3}$, respectively, which are based on the work
of Kouvaris (2008) \citep{Kouvaris08}. The solid curve shows the DM
density of 3.81 $\rm{GeV/cm^3}$ in our work taking the Bondi
accretion of DM particles into account.} \label{fig1}
\end{figure}

For convenience, we assume that the NS is made up of nonsuperfluid
neutrons, protons and electrons, and suppose that the density of the
star is uniform. In our specific calculations, we consider a
canonical NS of $M=1.4~M_{\odot}$ and $R=10~{\rm km}$, and impose
the initial interior temperature $T_0=10^{10}$ K at very early time
for the star. However, it is emphasized that the thermal evolution
of the star is very insensitive to the initial condition, which
affects the temperature only during the first years. Meanwhile, we
safely take the average velocity at infinity
$\bar{\upsilon}_\infty=220~\rm{km/s}$ of DM particles. It is worth
stressing that the DM density at infinity $\rho(r_\infty)=3.81
~\rm{GeV/cm^3}$ is adopted. It is the critical DM density whether
considering the Bondi accretion of DM by neutron stars or not.

The evolution of the surface temperature for a typical NS with
$M=1.4~M_{\odot}$ and $R=10~{\rm km}$ as a function of time is shown
in Fig. \ref{fig1}. For the ``standard cooling" scenario, the
cooling of a NS will undergo the neutrino emission and photon
emission. After $\sim10^6$ years, the surface temperature of the
star drops fast during the photo cooling era. Figure \ref{fig1}
shows the cases where the effect of DM annihilation is considered
for three different DM densities, based on the accretion mechanism
of DM particles discussed in the work of Kouvaris (2008)
\citep{Kouvaris08}. It can be seen from the figure that DM heating
can affect the surface temperature of the star significantly after
$\sim 10^7$ year as the temperature of the star drops. The plateau
of surface temperature will appear, once the equilibrium between the
heating of DM annihilation and the cooling of surface photon
emission has been reached. Furthermore, the temperature platform
depends on the DM density, the mass, and the radius of the star.
Obviously, the temperature platform is higher and appears earlier
for the case with higher DM density. However, it is not optimistic
to detect the surface temperature as low as $\sim 10^4$ K for the
case with the DM density of 30 $\rm{GeV/cm^3}$, which would possibly
be a signature of DM annihilation. Encouragingly, the surface
temperature platform can reach up to $\sim 1.12\times10^5$ K, if
Bondi accretion will play the leading role in the accretions of DM
particles by neutron stars. The temperature platform is about one
order of magnitude higher and appears earlier ($\sim 10^{6.5}$ year)
than that of Kouvaris (2008), because the Bondi accretion rate of DM
particles is well above that of Kouvaris (2008). As it can be
deduced from Eq. (\ref{accretion rate 2}), the Bondi accretion rate
of DM particles can reach up to $1.31\times10^8~ \rm{g/s}$, which is
about six orders of magnitude higher than that of Kouvaris (2008)
(see Eq. (18) in Ref. \citep{Kouvaris08}). It is possible to explain
the high temperature behaviors of old neutron stars.
\\
\section{CONCLUSIONS} \label{conclusions}

We have shown the effects of DM heating on the surface temperature
of a canonical NS ($M=1.4~M_{\odot}$, $R=10~{\rm km}$) for two
different accretion mechanisms of DM particles. DM heating will play
a leading role at the later time of NS, as the temperature of the
star drops. Once the heating of DM annihilation equilibrates the
cooling of surface photon emission, as a result the temperature of
the star will remain flat as a function of time. The temperature
plateau depends largely on the DM density, the mass, and the radius
of NS. The Bondi accretion of DM by neutron stars is dominant rather
than the accretion mechanism of Kouvaris (2008) \citep{Kouvaris08},
if the DM density is higher than $\sim3.81~\rm GeV/cm^3$. It is
obvious that the surface temperature platform appears at $\sim
10^{6.5}$ year, which is earlier than that of Kouvaris (2008) with
three different DM densities. In particular, the surface temperature
platform can arrive $\sim 1.12\times10^5$ K for the DM density of
$3.81~\rm GeV/cm^3$, which is about one order of magnitude higher
than the case of Kouvaris (2008) with DM density of $30~\rm
GeV/cm^3$. More importantly, it maybe explain the high temperature
behaviors of old neutron stars.

The model we adopted is based on uniform stellar configuration.
However, it is commonly considered that the neutron stars can be
approximated as the uniform case, because the results will be
unchanged in the order of magnitude when considering the realistic
equations of states. In addition, the effect of stellar rotation is
disregard, which can lead to rotochemical heating due to the
derivative from beta equilibrium with the spin-down of the star.
This mechanism could possibly be present for old neutron stars.

\section*{\uppercase {acknowledgments}}
The authors would like to thank the referee very much for helpful
comments and Yunwei Yu for useful discussions, which have
significantly improved our work. This work is supported by
Scientific Research Project Fund of Hubei Provincial Department of
Education (Grant No. Q20161604).


\begin{thebibliography}{}

\bibitem[Ade et al., 2016]{Ade16}P. A. R. Ade {\it et al.} (Planck Collaboration), Astron. Astrophys. 594, 13 (2016).

\bibitem[Bi et al., 2013]{Bi13}X. J. Bi, P. F. Yin, and Q. Yuan, Front. Phys. 8, 794 (2013).

\bibitem[Klasen et al., 2015]{Klasen15}M. Klasen, M. Pohl, and G. Sigl, Prog. Part. Nucl. Phys. 85, 1 (2015).

\bibitem[Goodman and Witten, 1985]{Goodman85}M. W. Goodman and E. Witten, Phys. Rev. D 31, 3059 (1985).

\bibitem[Angloher et al., 2016]{Angloher16}G. Angloher {\it et al.} (CRESST-II Collaboration), Eur. Phys. J. C 76, 25 (2016).

\bibitem[Agnese et al., 2016]{Agnese16}R. Agnese {\it et al.} (SuperCDMS Collaboration), Phys. Rev. Lett. 116, 071301 (2016).

\bibitem[Agnes et al., 2016]{Agnes16}P. Agnes {\it et al.} (DarkSide Collaboration), Phys. Rev. D 93, 081101 (2016).

\bibitem[Akerib et al., 2017]{Akerib17}D. S. Akerib {\it et al.} (LUX Collaboration), Phys. Rev. Lett. 118, 021303 (2017).

\bibitem[Aprile et al., 2016]{Aprile16}E. Aprile {\it et al.} (XENON100 Collaboration), Phys. Rev. D 94, 122001 (2016).

\bibitem[Tan et al., 2016]{Tan16}A. Tan {\it et al.} (PandaX-II Collaboration), Phys. Rev. Lett. 117, 121303 (2016).

\bibitem[Bertone et al., 2005]{Bertone05}G. Bertone, D. Hooper, and J. Silk, Phys. Rep. 405, 279 (2005).

\bibitem[Rott, 2013]{Rott13}C. Rott, Nucl. Phys. B, Proc. Suppl. 235, 413 (2013).

\bibitem[Conrad, 2014]{Conrad14}J. Conrad, arXiv:1411.1925.

\bibitem[Aad et al., 2015]{Aad15}G. Aad {\it et al.} (ATLAS Collaboration), Eur. Phys. J. C 75, 299 (2015).

\bibitem[Khachatryan et al., 2016]{Khachatryan16}V. Khachatryan {\it et al.} (CMS Collaboration), Phys. Lett. B 755, 102 (2016).

\bibitem[Baer et al., 2015]{Baer15}H. Baer, K.-Y. Choi, J. E. Kim, and L. Roszkowski, Phys. Rep. 555, 1 (2015).

\bibitem[Hooper and Profumo, 2007]{Hooper07}D. Hooper and S. Profumo, Phys. Rep. 453, 29 (2007).

\bibitem[Jungman et al., 1996]{Jungman96}G. Jungman, M. Kamionkowski, and K. Griest, Phys. Rep. 267, 195 (1996).

\bibitem[Weinberg, 1978]{Weinberg78}S. Weinberg, Phys. Rev. Lett. 40, 223 (1978).

\bibitem[Wilczek, 1978]{Wilczek78}F. Wilczek, Phys. Rev. Lett. 40, 279 (1978).

\bibitem[Covi et al., 2001]{Covi01}L. Covi, H. B. Kim, J. E. Kim, and L. Roszkowski, J. High Energy Phys. 05, 033 (2001).

\bibitem[Pagels and Primack, 1982]{Pagels82}H. Pagels and J. R. Primack, Phys. Rev. Lett. 48, 223 (1982).

\bibitem[Goldberg and Hall, 1986]{Goldberg86}H. Goldberg and L. J. Hall, Phys. Lett. B 174, 151 (1986).

\bibitem[Carlson et al., 1992]{Carlson92}E. D. Carlson, M. E. Machacek, and L. J. Hall, Astrophys. J. 398, 43 (1992).

\bibitem[Press and Spergel, 1985]{Press85}W. H. Press and D. N. Spergel, Astrophys. J. 296, 679 (1985).

\bibitem[Gould, 1987]{Gould87}A. Gould, Astrophys. J. 321, 571 (1987).

\bibitem[Gould, 1988]{Gould88}A. Gould, Astrophys. J. 328, 919 (1988).

\bibitem[Goldman and Nussinov, 1989]{Goldman89}I. Goldman and S. Nussinov, Phys. Rev. D 40, 3221 (1989).

\bibitem[Kouvaris, 2008]{Kouvaris08}C. Kouvaris, Phys. Rev. D 77, 023006 (2008).

\bibitem[Bertone and Fairbairn, 2008]{Bertone08}G. Bertone and M. Fairbairn, Phys. Rev. D 77, 043515 (2008).

\bibitem[Kouvaris and Tinyakov, 2010]{Kouvaris10}C. Kouvaris and P. Tinyakov, Phys. Rev. D 82, 063531 (2010).

\bibitem[de Lavallaz and Fairbairn, 2010]{de Lavallaz10}A. de Lavallaz and M. Fairbairn, Phys. Rev. D 81, 123521 (2010).

\bibitem[Fan et al., 2011]{Fan11}Y. Z. Fan, R. Z. Yang, and J. Chang, Phys. Rev. D 84, 103510 (2011).

\bibitem[Huang et al., 2014]{Huang14}X. Huang, W. Wang, and X. P. Zheng, Sci. China Phys. Mech. Astron. 57, 791 (2014).

\bibitem[Huang et al., 2015]{Huang15}X. Huang, X. P. Zheng, W. H. Wang, and S. Z. Li, Phys. Rev. D 91, 123513 (2015).

\bibitem[Kouvaris and Perez-Garcia, 2014]{Kouvaris14}C. Kouvaris and M. A. Perez-Garcia, Phys. Rev. D 89, 103539 (2014).

\bibitem[Bondi, 1952]{Bondi52}H. Bondi, Mon. Not. R. Astron. Soc. 112, 195 (1952).

\bibitem[Kato et al., 1998]{Kato98}S. Kato, J. Fukue, and S. Mineshige, {\it Black-Hole Accretion Disks} (Kyoto University Press, Kyoto, 1998).

\bibitem[Edgar, 2004]{Edgar04}R. Edgar, New Astron. Rev. 48, 843 (2004).

\bibitem[David and Paul, 2000]{David00}N. S. David and J. S. Paul, Phys. Rev. Lett. 84, 3760 (2000).

\bibitem[Kaplinghat et al., 2014]{Kaplinghat14}M. Kaplinghat, S. Tulin, and H. B. Yu, Phys. Rev. D 89, 035009 (2014).

\bibitem[Kamada et al., 2016]{Kamada16}A. Kamada, M. Kaplinghat, A. B. Pace, and H. B. Yu, arXiv:1611.02716.

\bibitem[Robertson et al., 2017]{Robertson17}A. Robertson, R. Massey, and V. Eke, Mon. Not. R. Astron. Soc. 465, 569 (2017).

\bibitem[Schaab et al., 1999]{Schaab99}Ch. Schaab, A. Sedrakian, F. Weber, and M. K. Weigel, Astron. Astrophys. 346, 465 (1999).

\bibitem[Gonzalez and Reisenegger, 2010]{Gonzalez10}D. Gonzalez and A. Reisenegger, Astron. Astrophys. 522, 16 (2010).

\bibitem[Shibazaki and Lamb, 1989]{Shibazaki89}N. Shibazaki and F. K. Lamb, Astrophys. J. 346, 808 (1989).

\bibitem[Larson and Link, 1999]{Larson99}M. B. Larson and B. Link, Astrophys. J. 521, 271 (1999).

\bibitem[Cheng et al., 1992]{Cheng92}K. S. Cheng, W. Y. Chau, J. L. Zhang, and H. F. Chau, Astrophys. J. 396, 135 (1992).

\bibitem[Goldreich and Reisenegger, 1992]{Goldreich92}P. Goldreich and A. Reisenegger, Astrophys. J. 395, 250 (1992).

\bibitem[Thompson and Duncan, 1996]{Thompson96}C. Thompson and R. C. Duncan, Astrophys. J. 473, 322 (1996).

\bibitem[Heyl and Kulkarni, 1998]{Heyl98}J. S. Heyl and S. R. Kulkarni, Astrophys. J. Lett. 506, L61 (1998).

\bibitem[Pons et al., 2007]{Pons07}J. Pons, B. Link, J. Miralles and U. Geppert, Phys. Rev. Lett. 98, 071101 (2007).

\bibitem[Reisenegger, 1995]{Reisenegger95}A. Reisenegger, Astrophys. J. 442, 749 (1995).

\bibitem[Reisenegger, 1997]{Reisenegger97}A. Reisenegger, Astrophys. J. 485, 313 (1997).

\bibitem[Fern\'andez and Reisenegger, 2005]{Fer05}R. Fern\'andez and A. Reisenegger, Astrophys. J. 625, 291 (2005).

\bibitem[Yuan and Zhang, 1999]{Yuan99}Y. F. Yuan and J. L. Zhang, Astron. Astrophys. 344, 371 (1999).

\bibitem[Yu and Zheng, 2006]{Yu06}Y. W. Yu and X. P. Zheng, Astron. Astrophys. 450, 1071 (2006).

\bibitem[Alford et al., 2005]{Alford05}M. Alford, P. Jotwani, C. Kouvaris, J. Kundu, and K. Rajagopal, Phys. Rev. D 71, 114011 (2005).

\bibitem[Shapiro and Teukolsky, 1983]{Shapiro83}S. L. Shapiro and S. A. Teukolsky, {\it Black Holes, White Dwarfs, and Neutron Stars} (Wiley, New York, 1983), p. 645.

\bibitem[Gudmundsson et al., 1983]{Gudmundsson83}E. H. Gudmundsson, C. J. Pethick, and R. I. Epstein, Astrophys. J. 272, 286 (1983).

\bibitem[Yang, 1992]{Yang92}L. T. Yang, {\it Hydromechanics and Theories of Accretion Disks} (Science Press, Beijing, 1992).

\end{thebibliography}
\end{document}